# Quantum critical Bose gas in the two-dimensional limit in the honeycomb antiferromagnet YbCl$_3$ under magnetic fields

Yosuke Matsumoto[1], Simon Schnierer[1], Jan A. N. Bruin[1], Jürgen Nuss[1], Pascal Reiss[1], George Jackeli[1,2], Kentaro Kitagawa[3], and Hidenori Takagi[1,2,3]

[1]Max Planck Institute for Solid State Research, Heisenbergstrasse 1, 70569 Stuttgart, Germany

[2]Institute for Functional Matter and Quantum Technologies, University of Stuttgart, Pfaffenwaldring 57, 70569 Stuttgart, Germany

[3]Department of Physics, University of Tokyo, Bunkyo-ku, Hongo 7-3-1, Tokyo 113-0033, Japan

**Bose-Einstein condensation (BEC) [1,2] is a quantum phenomenon, where a macroscopic number of bosons occupy the lowest energy state and acquire coherence at low temperatures. It is realized not only in superfluid $^4$He [3] and dilute atomic gases [4,5], but also in quantum magnets [6–13]. In three-dimensional (3D) antiferromagnets, an XY-type long-range ordering (LRO) occurs near a magnetic-field-induced transition to a fully polarized state (FP) and has been successfully described as a BEC in the last few decades. An attractive extension of the BEC in 3D magnets is to make their two-dimensional (2D) analogue. For a strictly 2D system, it is known that BEC cannot take place due to the presence of a finite density of states at zero energy, and a Berezinskii-Kosterlitz-Thouless (BKT) transition may instead emerge. In a realistic quasi-2D magnet consisting of stacked 2D magnets, a small but finite interlayer coupling stabilizes marginal LRO and BEC, but such that 2D physics is still expected to dominate. A few systems were reported to show such 2D-limit BEC, but at very high magnetic fields that are difficult to access. The honeycomb $S = 1/2$ Heisenberg antiferromagnet YbCl$_3$ with an intra-layer coupling $J \sim 5$ K exhibits a transition to a FP state at a low in-plane magnetic field of $H_s = 5.93$ T. Here, we demonstrate that the LRO right below $H_s$ is a BEC but close to the 2D-limit, marginally stabilized by an extremely small interlayer coupling $J_\perp$. At the quantum critical point $H_s$, we capture 2D-limit quantum fluctuations as the formation of a highly mobile, interacting 2D Bose gas in the dilute limit. A much-reduced effective boson-boson repulsion $U_{\text{eff}}$ as compared with that of a prototypical 3D system indicates the presence of a logarithmic renormalization of interaction unique to 2D. The old candidate for a Kitaev quantum spin liquid, YbCl$_3$, is now established as an ideal arena for a quantum critical BEC in the 2D limit.**

XY-antiferromagnetism induced by an applied magnetic field $H$ provides a prominent example of BEC in quantum magnets, which has been established in a wide variety of magnets including spin-singlet dimers TlCuCl$_3$ [6,7], BaCuSi$_2$O$_6$ [8,9] and $S = 1$ magnet with a single ion anisotropy NiCl$_2$-4SC(NH$_2$)$_2$ [10]. Consider, for example, the case for Heisenberg antiferromagnet with a nearest neighbor coupling $J$ in a field $H$. $H$ along the $z$-direction polarizes the spins and causes their $z$-component $<S_z>$ to acquire a finite value. When $H$ is close to the saturation field $H_s$, i.e., near a quantum phase transition to a fully polarized (FP) state, the system effectively becomes an XY antiferromagnet with the remaining $x$- and $y$- components of the spins, $S_x$ and $S_y$. As $S_x$ and $S_y$ can be replaced with creation/annihilation operators of bosons, the system can be mapped onto an ensemble of interacting bosons with boson operators $b_k^+$ and $b_k$ in the momentum $k$ representation, an excitation energy $\varepsilon_k$ and an effective chemical potential $\mu_{\text{eff}}$ [14], which is described by the following Hamiltonian.

$$H = \Sigma_k (\varepsilon_k - \mu_{\text{eff}}) b_k^+ b_k \quad (1)$$

In this mapping, $\varepsilon_k$ is determined by the energy dispersion of a tight-binding band with a nearest neighbor hopping $t = J/2$ measured from the bottom of the band. The effective chemical potential

$$\mu_{\text{eff}} = g\mu_B(H_s - H) - 2U_{\text{eff}}<n>$$

consists of the bare chemical potential controlled by $H_s - H$ and an additional term representing the increase of the mean-field, one-particle energy due to a repulsive boson-boson interaction $U_{\text{eff}}$, which arises from the $JS_z^i S_z^j$ term in the original spin Hamiltonian and is on the order of $J$ in the bare form [6,15]. $<n>$ is the average number of (hole) bosons per site, which corresponds to the deviation of field-induced magnetization $<S_z>$ from the saturation value. The XY-ordering below $H_s$ can therefore be treated as a BEC of a low density of interacting bosons in the Hamiltonian. The FP state above $H_s$ has an excitation gap and zero boson density $<n> = 0$ at temperature $T = 0$, and can be regarded as the vacuum state of the bosons. The field-induced quantum phase transition at $H_s$ provides us with a unique opportunity to study a BEC quantum critical point (BEC-QCP).

The nature of the magnetic-field-induced BEC depends sensitively on the dimensionality of the system. In a three-dimensional (3D) magnet, BEC occurs simply as a long-range XY ordering. Quite distinct from the 3D case is a two-dimensional (2D) magnet. It is known that a strictly 2D Bose gas does not experience BEC at a finite temperature, due to the presence of a finite density of states at zero energy and the associated logarithmic divergence of the number integral in the $T = 0$ limit. In the original language of spin, a long-range XY ordering in a strictly 2D magnet is suppressed by strong fluctuations down to $T = 0$ [16]. Instead, a quasi-LRO emerges, i.e., the Berezinskii-Kosterlitz-Thouless (BKT) transition[17–19]. One therefore observes a BKT near $H_s$ in strictly 2D systems, instead of the BEC. The effective boson-boson interaction $U_{\text{eff}}$ in 2D is distinct from those in 3D and is subject to logarithmic renormalization as $U_{\text{eff}} = U_0/(-\ln<n>)$, where $U_0$ is the bare interaction, and suppressed appreciably around the QCP [20]. In reality, the bulk "2D magnet" that we investigate experimentally is not a strictly 2D magnet but a quasi-2D magnet, which comprises a stack of strictly 2D magnets with a small but finite inter-layer coupling ($J_\perp$). The inter-layer coupling should marginally push the system from a BKT transition to a 3D LRO and hence a BEC. Even in such a quasi-2D BEC state, however, 2D physics could be still relevant at temperatures above the temperature scale of $J_\perp$. 2D logarithmic renormalization of $U_{\text{eff}}$ for example may still be captured. Possible signatures of BKT physics were suggested theoretically[21] and experimentally[22–25] in quasi-2D magnets with intrinsic XY character.

It is tempting to explore such a distinct class of marginal BEC in the 2D limit and its BEC-QCP in quasi-2D antiferromagnets under magnetic fields, where we anticipate 2D logarithmic renormalization, 2D quantum and thermal fluctuations with possible signatures of BKT physics and an interplay with 3D physics arising from a minute interlayer coupling $J_\perp$. So far, despite the long history of BEC and BEC-QCP in antiferromagnets, such exploration for the 2D limit has been limited due to the lack of an appropriate model system. The quasi-2D dimer system BaCu$_2$Si$_2$O$_6$ shows a magnetic-field-induced BEC above the QCP where a finite magnetization emerges. The reduction of the effective interlayer coupling due to frustration was argued to play a vital role[9]. The linear decrease of the transition temperature $T_c$ as a function of the magnetic field around QCP, which is expected for a BEC in the 2D limit[26], suggests the presence of 2D quantum fluctuations. However, the frustration scenario was challenged by the later observation of ferromagnetic inter-layer coupling. The presence of two types of dimers does not allow for a straightforward interpretation of the critical behavior[27]. More importantly, signatures of "2D" critical fluctuations other than the linear scaling of transition temperature and underlying boson-boson interactions, however, have not yet been unveiled in contrast to the 3D case (*e.g.*, TlCuCl$_3$ [6] and NiCl$_2$-4SC(NH$_2$)$_2$ [28]), partly due to the very high magnetic field needed to reach the QCP. To capture the BEC criticality in the 2D limit experimentally, a material system with an easily accessible QCP is highly desired, where one can probe 2D quantum and thermal fluctuations and underlying interactions through other thermodynamic parameters in addition to $T_c$.

The pseudo-spin 1/2 Heisenberg antiferromagnet YbCl$_3$ may be an ideal system for the exploration of the quasi-2D BEC-QCP. The material has the 2D honeycomb-based structure shown in the inset of Fig. 1a [29] and was earlier suggested as a possible Kitaev magnet with anisotropic bond-dependent couplings[30]. Recent inelastic neutron scattering (INS) measurements, however, revealed that the system is a quasi-2D Heisenberg antiferromagnet with an almost isotropic in-plane nearest-neighbor coupling $J \approx 5$ K (= 0.42 meV) and a very small interlayer coupling $|J_\perp/J| \sim 3 \times 10^{-5}$ [31]. Recent quantum Monte Carlo (QMC) simulations on this system indicate a ferromagnetic interlayer coupling at least smaller than $|J_\perp/J| \sim 2 \times 10^{-3}$ [32], consistent with the estimate from the INS measurement [31]. $J_\perp$ is therefore extremely small, likely of the order of 0.1 mK, and at most 10 mK, rendering this system ideal to explore a BEC close to the 2D limit. Specific heat and neutron diffraction measurements indicate the occurrence of a 3D long-range Néel ordering at $T_N = 0.6$ K [30] with an ordered moment of ~1 $\mu_B$ [30, 31, 33], which is stabilized by the small $J_\perp$. AC susceptibility measurements indicate a magnetic-field-induced transition to a fully polarized state at $H_s = 6$ T and 9.5 T with the field applied in and out of plane, respectively [30], which we argue to be a quasi-2D BEC-QCP.

We therefore have explored the quasi-2D BEC-QCP in YbCl$_3$ with the in-plane magnetic field $H$ as a tuning parameter of the quantum phase transition. At the QCP, we identified clear signatures of BEC quantum critical fluctuations in the 2D limit, which manifest themselves as the formation of a highly mobile, correlated 2D Bose gas in the dilute limit, where the effective boson-boson interaction is an order of magnitude smaller than those of its 3D analogues due to the expected logarithmic renormalization of boson-boson interaction. The finite temperature transitions below the saturation field $H_s$ (*i.e.*, the QCP) can be described as a BEC induced by an extremely small interlayer coupling $J_\perp$ of ~ 0.1 mK.

**Phase diagram, Heisenberg-like to XY-like crossover and the QCP**

Single crystals of YbCl$_3$ used in this study were grown by a chemical vapor transport technique (see **Methods**). The magnetic field was always applied along the in-plane *a*-direction, perpendicular to one of the honeycomb bonds. The magnetization $M$ shown in Fig. 1a reaches the saturation moment $M_s = 1.72$ $\mu_B$/Yb around $H_s \sim 5.9$ T. The saturation field in the $T = 0$ limit, which marks the quantum phase transition, was estimated as $H_s = 5.93 \pm 0.01$ T from the crossing point of $dM/dH$ curves at 0.05 and 0.08 K (see Supplementary Information (**SI**)).

At zero field, the specific heat divided by temperature $C/T$ shown in Fig. 2b exhibits a broad peak from a short-range 2D antiferromagnetic correlation around 1.2 K, followed by a tiny but sharp peak from the long-range 3D Néel ordering at $T_c = 0.65$ K, fully consistent with the previous studies [30, 33]. Weak signature of the Néel order is also present in $M(T)$, which is more clearly visualized in the temperature derivative $dM/dT$ (Fig. 2e, f). Upon applying $H$, $T_c$ first increases until $H_p \sim 2$-3 T, then decreases to zero at the saturation field $H_s = 5.93$ T [30, 33]. This evolution is summarized in the phase diagram shown in Fig. 1c. The initial increase of $T_c$ indicates the suppression of fluctuations along the field direction and the crossover of symmetry from Heisenberg-like to XY-like, as discussed in a class of low-dimensional Heisenberg magnets [34,35]. The suppression of fluctuations can indeed be captured by the initial decrease of the isothermal entropy up to $H_p$ at temperatures below 0.8 K in Fig. 1b, as well as a negative slope $dM/dT|_H (= dS/dH|_T)$ in the same $H$- and $T$-range in Fig. 2a, e.

Above $H_p \sim$ 2-3 T, the system should have predominant XY character. Reflecting this change of symmetry, the anomalies at $T_c$ in $C/T$ and $M$ show qualitatively different behavior from that of the zero-field limit. In $C/T$, the small λ-like peak associated with LRO and the broad peak associated with short-range ordering (SRO) merge at higher fields into one sharp cusp-like peak (Fig. 2b). The weak anomaly in $dM/dT$ at low fields changes to a cusp-like anomaly near $H_p$ (Fig. 2e, f). As $H$ increases further toward $H_s$, the LRO with XY-character is suppressed due to the reduced $S_x$ and $S_y$ degrees of freedom and $T_c$ decreases to zero (Fig.1c). In this field region near $H_s$, the LRO with XY character is described as a quasi-2D BEC induced by interlayer coupling, as we will discuss below. In the language of bosons, the suppression of the quasi-2D BEC upon approaching the QCP at $H_s$ can be described by a decrease of the boson density to zero.

Above $H_s$, we observe thermally activated behaviors of $C$ and $(M_s-M)/M_s$ at low temperatures (see Supplementary Fig. S4a, b), indicating the emergence of a gap in the magnetic (boson) excitations in the FP state. The extracted activation energy Δ, as shown in Fig. 1c and Fig. S4c, increases linearly from $H_s$ as roughly $g\mu_B(H_s-H)$, where $g \sim 3.67$ is the g-factor for $S = 1/2$ pseudo-spins (see **SI**). In the language of bosons, $g\mu_B(H_s-H)$ corresponds to the energy between the bare chemical potential $\mu$ and the bottom of the band and sets the excitation gap above $H_s$. These behaviors above $H_p$ indicate that the honeycomb antiferromagnet YbCl$_3$ under magnetic field is an excellent arena to explore a quasi-2D BEC and the associated BEC-QCP.

**2D-limit BEC critical fluctuations at the QCP**

At the saturation field $H_s$, *i.e.*, the QCP, we indeed find evidence for quantum fluctuations predicted for a BEC-QCP in the 2D limit. The critical behaviors of $C(T)$, $M(T)$ and $T_c(H)$ at a field-induced BEC-QCP in $d$ dimensions are predicted to be $C \propto T^{d/2}$, $M_s-M \propto T^{d/2}$ and $T_c(H) \propto (H_s – H)^{2/d}$ [11, 26, 36, 37]. In the 3D BEC systems such as TlCuCl3[6] and NiCl$_2$-4SC(NH$_2$)$_2$[28], the critical exponents with $d = 3$ was firmly established at the BEC-QCP. As summarized in Fig. 2b, c and 3a, b, in stark contrast to 3D model systems, all three parameters of YbCl$_3$, $C$, $M$ and $T_c$, closely follow the expected critical behavior in the 2D ($d = 2$) limit at $H_s$. $C$ is linear in $T$ with a coefficient $\gamma = C/T \sim$ 1 J/K$^2$ Yb-mol over a wide $T$ range below ~1.2 K. $M$ decreases linearly with $T$ from the saturation moment as $M = M_s – M_s (T/T_0)$ with $T_0 =$ 11 K. In the language of bosons, the boson density $<n> \equiv (M_s –M)/M_s = T/T_0$ goes to zero at $T = 0$. $T_c$ scales linearly with $H_s – H$ near the critical point $H_s$. These quantum critical behaviors can be utilized as markers for quantum fluctuations. The power exponent of $C(T)$, $\alpha(T) \equiv d \ln C/d \ln T$ is indicated as a color contour map on the $H$-$T$ phase diagram in Fig. 1c. The red region for $\alpha \sim 1$, which represents BEC criticality in the 2D limit, spreads like a fan from the QCP at $H_s$. The contour map of $dM/dT$ yields essentially the same quantum critical region (See **SI**). The fan-like spread is indeed the canonical behavior of a quantum critical regime around the QCP, which confirms that the critical exponents, which are consistent with a BEC-QCP in the 2D limit, arise from quantum critical fluctuations. The 2D quantum critical behavior is observed at least down to our lowest temperature of measurements, $T =$ 50 mK. We note that 50 mK is only 1% of $J =$ 5 K, but still orders of magnitude higher than the energy scale of $J_\perp$, which is on the order of 0.1 mK based on previous neutron scattering studies and our analysis below. A substantial part of the 3D critical behavior originating from the small $J_\perp$ is very likely hidden in the low-temperature limit below experimentally accessible temperatures.

**Formation of an interacting 2D Bose gas in the dilute limit at the QCP**

The quasi-2D BEC-QCP lies at the limit of zero boson density, where the system hosts a dilute and therefore weakly interacting boson gas produced by thermal excitations at a finite temperature. Let us consider a 2D boson gas with a constant density of states $D(E) = D$ and an effective chemical potential $\mu_{\text{eff}} = g\mu_B(H_s-H) − 2U_{\text{eff}}<n>$, as shown in Fig. 3c. For a tight-binding model on the 2D honeycomb lattice with nearest-neighbor hopping $t = J/2$, $D = \sqrt{3}/2\pi J$ at the bottom of the band (energy $E = 0$) (see **SI**). From the number integral with the Bose function, $\langle n \rangle$ and $\mu_{\text{eff}}$ are related by the following equation:

$$\exp(-\langle n \rangle / D k_B T) + \exp(\mu_{\text{eff}}/k_B T) = 1 \quad (2)$$

At the QCP with $H = H_s$, $\mu_{\text{eff}} = -2U_{\text{eff}}\langle n \rangle$. Equation (2) then requires $\langle n \rangle$ and hence $\mu_{\text{eff}}$ to be linear in $T$, in accord with the BEC critical behavior of $\langle n \rangle$ in the 2D limit. The experimentally obtained $T$-linear boson density from $M$ in Fig. 3a, $\langle n \rangle = T/T_0$ ($T_0 = 11$ K $= 2.2J$), yields $U_{\text{eff}} = 1.2$ K $= 0.24J$ from Eq. (2). It is known that the free 2D Bose gas ($U_{\text{eff}} = 0$) with zero chemical potential has a $T$-linear $C(T)$ at low temperatures with a linear coefficient $\gamma = (\pi^2/3)k_B^2 D$, the same expression as that of a free Fermi gas. With $D = \sqrt{3}/2\pi J$ and $J = 5$ K, the free boson $\gamma$ is 1.5 J/molK$^2$. We find that the $T$-linear negative shift of the chemical potential from zero, $\mu_{\text{eff}} = -2U_{\text{eff}}\langle n \rangle = -0.21 k_B T$, reduces the $\gamma$ value from the free boson value to 0.99 J/molK$^2$, in excellent agreement with the experimental data (see SI). These results firmly justify the estimate of $U_{\text{eff}}$ and $\mu_{\text{eff}}$, and more importantly, the 2D Bose gas description. The BEC quantum criticality in the 2D limit manifests itself as the formation of an interacting 2D Bose gas at the zero-density limit.

$U_{\text{eff}} = 0.24J$ is an order of magnitude smaller than those estimated for prototypical 3D BEC systems, $5J$ for TlCuCl$_3$ [38,39] and $3J$ for NiCl$_2$-4SC(NH$_2$)$_2$ [40]. We argue that this represents the logarithmic renormalization of boson-boson scattering $U_0$ unique to 2D, $U_{\text{eff}} \sim -U_0/\ln\langle n \rangle$ [20], and mirrors the 2D character of quantum critical Bose gas in YbCl$_3$. The 2D renormalization alone would bring $U_{\text{eff}}$ to zero at the quantum critical point due to the logarithmic divergence. We argue that $U_{\text{eff}}$ stays at a finite value $\sim 0.24J$ even at the QCP here due to the weak three-dimensionality associated with the interlayer coupling $J_\perp$, which suppresses the logarithmic singularity at the bottom of the band, as we discuss later. The cut-off of logarithmic divergence is roughly estimated as $U_c \sim -U_0/\ln(J_\perp/J)$, which suggests $-\ln(J_\perp/J) \sim 10$ for YbCl$_3$.

**BEC due to a finite interlayer coupling $J_\perp$**

Lowering the applied magnetic field below $H_s$, which corresponds to boson doping, clear anomalies indicative of a phase transition emerge in the $T$-dependent $C(T)$ and $M(T)$ at $T_c(H)$, as shown in Fig.2. We found that $T_c(H)$ can be quantitatively described as a BEC of 3D system by introducing an extremely small interlayer coupling $J_\perp$ to purely 2D band, which implies that the transitions are a long-range magnetic ordering stabilized by the interlayer coupling rather than the BKT transition for 2D. The presence of a small interlayer coupling $J_\perp$ rounds the bottom of the purely 2D band and produces a corresponding $\sqrt{E}$-dependent density of states, as expected in 3D, within the extremely narrow energy range set by $J_\perp$ (Fig. 3c). The continuously vanishing density of states at $E = 0$ prevents a logarithmic divergence of the number integral and hence gives rise to a BEC at a finite temperature. We approximate the rounding of the constant 2D density of states $D(E) = D$ near the bottom of the band by introducing an energy cut-off $E_c$, below which $D(E)$ is replaced with the 3D density of states, $\sqrt{E}/E_c$. $E_c$ is of the order of $J_\perp$, roughly 2-3 $J_\perp$. By setting $\mu_{\text{eff}} = 0$ in the number integral, a BEC occurs at $\langle n_c \rangle$ and $T_{\text{BEC}}$ satisfying

$$\langle n_c \rangle \approx D k_B T_{\text{BEC}} (-\ln(k_B T_{\text{BEC}}/E_c) + 2). \quad (3)$$

The first and the second terms in Eq. (3) come from the 2D boson density above $E_c$ and the small 3D contribution below $E_c$ respectively. In Fig. 3a, we overlay Eq. (3) with $D = \sqrt{3}/2\pi J$, $J = 5$ K and $E_c = 0.2$ mK ($=2J_\perp$ for $J_\perp/J = 2\times10^{-5}$) as a solid line. The predicted $\langle n_c \rangle$ and $T_{\text{BEC}}$ reasonably reproduce the experimentally observed $T_c$ and $\langle n_c \rangle$ (arrows in Fig. 3a). This close agreement clearly indicates that the magnetic ordering near the QCP is described as a BEC stabilized by a very small interlayer coupling $J_\perp$ of the order of $10^{-5} J$. The extremely small $J_\perp$ compared to $J$ implies that YbCl$_3$ is very close to the 2D limit. We note that the estimate of interlayer coupling is fully consistent with that in a previous neutron scattering study [31].

At the BEC, the chemical potential $\mu_{\text{eff}} = g\mu_B(H_s-H) - 2U_{\text{eff}}\langle n_c\rangle = 0$. This gives an estimate of the effective interaction $U_{\text{eff}} = g\mu_B(H_s-H)/2\langle n_c\rangle$ for $H < H_s$, which is plotted as a function of $\langle n_c\rangle$ in Fig. 3d. With $\langle n_c\rangle \to 0$, $U_{\text{eff}}$ only weakly decreases to $U_{\text{eff}} \sim 0.2\,J$, which is estimated from the analysis of a 2D quantum critical Bose gas at $H=H_s$. The decrease can be fitted reasonably by $U_{\text{eff}} = -U_0(1/\ln\langle n_c\rangle + 1/\ln(J_\perp/J))$ with $U_0 = 6$ K $= 1.2J$ and $J_\perp/J = 2 \times 10^{-5}$ with 2D logarithmic renormalization and 3D cut-off, as seen by the dotted line in Fig. 3d. $U_0$ is smaller than but reasonably close to the $U_{\text{eff}} = 3\text{-}5J$ estimated for the canonical 3D BEC systems. Note that $U_{\text{eff}} = -U_0/\ln\langle n_c\rangle$ go to 0 with $\langle n_c\rangle \to 0$ but stays a finite $U_{\text{eff}}$ of $\sim 0.2\,J$ at $\langle n_c\rangle \sim 0$, which we discussed as a cut-off by the interlayer coupling $J_\perp$. These observations firmly establish the presence of 2D logarithmic renormalization, one of the hallmarks of 2D physics.

**Thermal fluctuations in the 2D limit**

Reflecting the proximity of the system to the 2D limit, clear signatures of 2D thermal fluctuations are observed above $T_c$ in the specific heat data. In the specific heat power $\alpha(T)$-map in Fig. 1c, the red region of 2D quantum critical behavior with $\alpha \sim 1$ crosses over to a white region with $\alpha < 1$ with lowering temperature. The white area that extends down to $T_c$ corresponds to the accelerated increase of $C(T)/T$ from a $T$-independent behavior upon approaching $T_c$ in Fig. 2b, which marks the region of thermal fluctuations. It spreads over a wide range of temperatures, from as high as $\sim 2T_c$ (dotted line in Fig. 1c) down to $T_c$, which points to the 2D character of the thermal fluctuations. Below $T_c$, we see a decrease of $C(T)/T$ and a corresponding positive power $\alpha$ that decreases eventually to $\sim 2$. As the lowest temperature $\sim 100$ mK is still higher than $0.5T_c$ in the critical region near $H_s$ ($> 5.5$ T), it is difficult to extract the low-$T$ limit of $\alpha$ and to discuss the dimensionality of fluctuations below $T_c$.

Since the system is essentially an XY-magnet near the QCP and in the 2D limit, the 2D fluctuations observed above and below $T_c$ may carry certain characteristics of a BKT transition for the strictly 2D case. It is tempting to infer here that the hypothetical BKT transition temperature $T_{\text{BKT}}$ in the $J_\perp = 0$ limit is very likely close to the observed BEC transition temperature $T_c$. Theoretically, it was shown for classical spins that the long-range ordering temperature $T_c$ for a small $J_\perp = 0$ is only slightly above $T_{\text{BKT}}$[41]. QMC simulations of a $S = 1/2$ Heisenberg antiferromagnet on a purely 2D square lattice (not honeycomb lattice) under magnetic fields near the saturation field $H_s$[42] give an estimate of the BKT transition temperature $T_{\text{BKT}}/J \sim 1\text{-}H/H_s$, which is indeed reasonably close to the experimentally observed BEC transition temperatures $T_c$ in Fig. 1c.

**High mobility of 2D Bose gas at the QCP**

The interacting 2D Bose gas at the QCP is highly mobile at low temperatures, very likely due to the reduced boson-boson interactions in the dilute and 2D limit, which shows up as a singular enhancement of thermal conductivity $\kappa(T)$ at $H_s$. We plot the normalized differential thermal conductivity $\Delta\kappa/\kappa(11.9\,\text{T}) \equiv [\kappa(T, H) - \kappa(T, 11.9\,\text{T})]/\kappa(T, 11.9\,\text{T})$ as a color contour map on the $H$-$T$ phase diagram in Fig. 4a, where the positive contribution indicates the excess thermal conductivity originating from the magnetic heat carriers. (see **SI** for details of $\kappa(T)$.) At the highest measured field of 11.9 T, the heat flow carried by magnetic excitations and the scattering of phonons by magnetic excitations are negligibly small below 2 K as the magnetic excitation gap is well developed ($\gg 2$K). $\kappa(T)$ at 11.9 T is therefore a reference for the maximum phonon thermal conductivity in the absence of scattering by magnetic excitations. A positive $\Delta\kappa$ indicates the presence of additional contributions other than the phonon contribution, naturally those from magnetic excitations. A singular positive $\Delta\kappa$ emerges up to $\sim 0.5$ K as a vertical red spike with a width of $\sim 0.4$ T around $H_s$ in Fig. 4a, indicating that the magnetic contribution of thermal conductivity at low temperatures peaks sharply at $H_s$ independent of $T$. This can be confirmed in the isotherm in Fig. 4b.

The thermal conductivity of 2D magnetic excitations is expressed as $\kappa_{mag} = (1/2)C_{mag}\langle v_{mag}\rangle l_{mag}$, where $C_{mag}$, $v_{mag}$ and $l_{mag}$ are the specific heat, the velocity and the mean free path of magnetic excitations. $C_{mag}(H) \sim C(H)$ at low temperatures shows a peak at the magnetic transition field $H_c(T)$ as in Fig. 4c, which moves away to a lower field from $H_s$ with increasing $T$ and is distinct from the position of the $\Delta\kappa$ peak always at $H_s$. The singular enhancement of magnetic thermal conductivity $\Delta\kappa$ should therefore be dominated by the enhancement of $\langle v_{mag}\rangle l_{mag}$ at $H_s$. As $\langle v_{mag}\rangle \sim \sqrt{2mk_BT}$ ($m$ : boson mass) does not depend on $H$ strongly around $H_s$, $l_{mag}$ must be enhanced drastically in very a narrow region near the QCP. We note that, in the red spike-region of the rapid enhancement of $\Delta\kappa$ in Fig. 4a, the number of bosons $\langle n\rangle$ is less than 0.1. We argue that the low boson density $\langle n\rangle$ around $H_s$ reduces the dominant boson-boson scattering represented by $2U_{eff}\langle n\rangle$ and makes the quantum critical 2D Bose gas highly mobile, which drastically enhances $\Delta\kappa$. The 2D logarithmic suppression of boson-boson scattering $U_{eff}$ may further enhance $\Delta\kappa$ around the QCP. An increase of $\kappa$ around QCP was observed also in the 3D magnetic BEC system, $NiCl_2\text{-}4SC(NH_2)_2$ at very low temperatures[43]. The $\kappa$ peak as a function $H$ around the QCP, however, is appreciably broader than the present 2D case, if normalized by the field scale of $H_s$, and appears to be closely correlated with the $C(H)$ peak. In this 3D analogue, the dominant scattering of bosons is indeed ascribed to the static defects[43] rather than boson-boson interactions, in contrast to the present 2D case.

In summary, we identified a 2D-limit BEC quantum criticality in the honeycomb quasi-2D Heisenberg antiferromagnet $YbCl_3$ under magnetic fields around the saturation field $H_s$, where a magnetic-field-induced quantum phase transition to the fully polarized state takes place. At $H_s$, the QCP, the system behaves as a highly mobile dilute 2D gas of bosons in the density $\langle n\rangle = 0$ limit with the critical exponents of specific heat $C(T)$ and magnetization $M(T)$ predicted for the BEC-QCP in the 2D limit. Lowering the magnetic field to $H < H_s$, an extremely weak interlayer coupling $J_\perp \sim 10^{-5} J$ marginally stabilizes a 3D LRO below $T_c$, which can de described quantitatively as a BEC. Reflecting the 2D-limit nature, 2D quantum and thermal fluctuations are captured clearly above $T_c$. A small boson-boson interaction $U_{eff}$ of $\sim 0.2 J$, one order of magnitude smaller than those in 3D analogues, is observed at the QCP, which increases only weakly with lowering $H$ from the QCP, namely boson-doping. We argue that the drastic suppression and the weak $H$-dependence can be quantitatively described as the logarithmic renormalization of the bare boson-boson interaction unique to 2D. $YbCl_3$ is an ideal arena to explore the physics of 2D interacting hard-core bosons.

**Methods**

**Single crystals.** Single-crystalline samples of $YbCl_3$ used in this study, transparent and with a thin plate-like shape, were synthesized by a self-chemical-vapor-transport method. Polycrystalline anhydrous $YbCl_3$ (Sigma-Aldrich 99.99%) was used as a starting material and sealed under vacuum in a long quartz tube with inner diameter of 15 mm. The starting material was placed at the hot end of the tube, which was heated to 900 °C and subsequently cooled down to 750 °C at a rate of 1 °C/hour. The temperature difference between the hot and the cold ends of tube was approximately 100 °C during the growth. The negligibly small paramagnetic responses from the impurities in $M$ and the large boundary-limited $\kappa$ confirm the high quality of the single crystals. The crystallographic axis was checked by single-crystal X-ray diffraction. The magnetic field was applied always along the in-plane $a$-axis.

**Magnetization measurements.** The magnetization $M$ below 3 K was measured with a home-made Faraday magnetometer installed in a $^3$He-$^4$He dilution refrigerator. The measurements were conducted on a few pieces of crystals put together, aligned in the same direction, with a total mass of $\sim 0.3$ mg. These were covered with dried Apiezon-N grease to protect the crystals from oxidation. The oxidation of samples could be checked by the presence a paramagnetic response in the magnetization curve. We used data only from crystals with negligibly small traces of such response. The absolute value of $M$ was determined by calibrating the magnitude of the field-dependent Faraday signals at 2 K with previous data taken at 1.8 K[30]. The 0.2 K difference between the two measurements gives an error in the calibration up to $\sim 1\%$, which does not influence the conclusion of this work. We further checked that the calibrated data are consistent with those measured at high temperatures $T \geq 2$ K by a commercial setup (Quantum Design, PPMS, VSM option).

**Specific heat measurements.** The specific heat $C$ was measured by a relaxation calorimetry[44] for aligned crystals with a total mass $\sim 0.07$ mg, which were covered with Apiezon-N grease to avoid oxidation. The total mass was determined by matching the $C$ of single crystals with that of a polycrystalline sample at zero-field and below $\sim 1$ K, where the contribution from the grease can be reasonably neglected. The grease contribution was determined as a difference from the polycrystalline sample at higher $T$. This amounts to 15% of the total heat capacity at 2 K and shows roughly $T^2$-dependence, which can be ascribed to the contribution from amorphous phonons of the grease. We subtracted this from all the data including those under fields.

**Thermal conductivity measurements.** $\kappa$ was measured with a conventional steady state method using a $^3$He-$^4$He dilution refrigerator in a temperature range of 0.1-3 K. The sample with a dimension of $\sim 1$ mm $\times \sim 2$ mm $\times \sim 20$ µm$^t$ was mounted onto a homemade cell. The cell was sealed in a glove box under argon atmosphere, which was then evacuated in the cryostat through a simple pop-up valve mechanism. Two more samples with similar dimensions were measured using a $^3$He cryostat and the results were well reproduced in the $T$-range of overlap (0.3 K- 3 K).

### Acknowledgement

We acknowledge C. D. Batista, Y. Kato and D. Huang for stimulating discussions and K. Pflaum and M. Dueller for experimental support. This work was partly supported by the Alexander von Humboldt foundation and the Japan Society for the Promotion of Science (JSPS) KAKAENHI (Numbers JP22H01180 and JP17H01140) and by the Max-Planck-UBC-UTokyo Center for Quantum Materials and the National Science Foundation under Grant No. NSF PHY-1748958.

### Author contributions

Y.M, K.K and H.T conceived the research. K.K synthesized the single crystals. Y.M performed the magnetization and specific heat measurements. J.A.N.B, S.S and Y.M performed the thermal conductivity measurements. J.N performed the structural characterization of crystals. Y.M, J.A.N.B and H.T analyzed the data. G.J and H.T provided theoretical inputs. P.R and G.J verified the analysis. Y.M and H.T wrote the manuscript with inputs from all the authors.

### Competing interests

The authors declare no competing interests.

### Data availability

The data that support the plots within this paper and other findings in this study are available from the corresponding authors upon reasonable request.

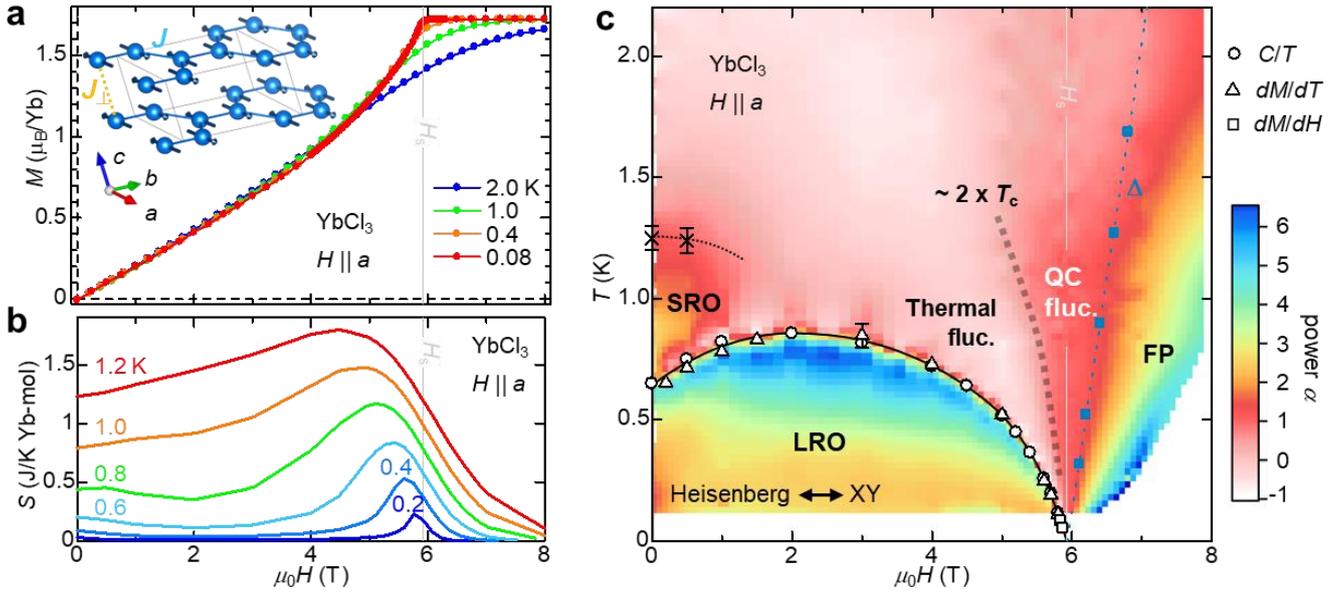

**Fig. 1 | Magnetic-field-induced quantum phase transition and *H-T* phase diagram for YbCl$_3$.** a) The magnetization *M*(*H*) curves at different temperatures. The Van-Vleck contribution has been subtracted (see SI for details). The inset shows the honeycomb layered structure of YbCl$_3$ (space group *C*12/*m*1) [29, 45] with an antiferromagnetic intralayer coupling *J* ≈ 5 K and a ferromagnetic interlayer coupling $J_\perp$ ~ 0.1 mK. b) The magnetic field dependence of entropy at different temperatures. c) The *H-T* phase diagram. The power α of the *T*-dependent specific heat *C* ($C \propto T^\alpha$ defined as $\alpha(T) \equiv d \ln C/d \ln T$) was evaluated by linear fits to ln *C* vs. ln *T* for every 4 data points and is indicated by the color. The open symbols indicate the $T_c$ of the long-range magnetic ordering, defined by the position of peaks in *dM/dT*, *C/T* and *dM/dH*. The crosses represent the locations of the broad SRO peaks in *C/T*. Filled squares indicate the gap size Δ determined by *C/T*. A linear fit to Δ is shown with dotted line above $H_s$. The grey broken line below $H_s$ indicates the onset of the 2D thermal fluctuations above $T_c$, which is determined as the onset temperature of a rapid increase of *C/T* upon cooling (the open arrows in Fig. 2b). Error bars represent one standard deviation.

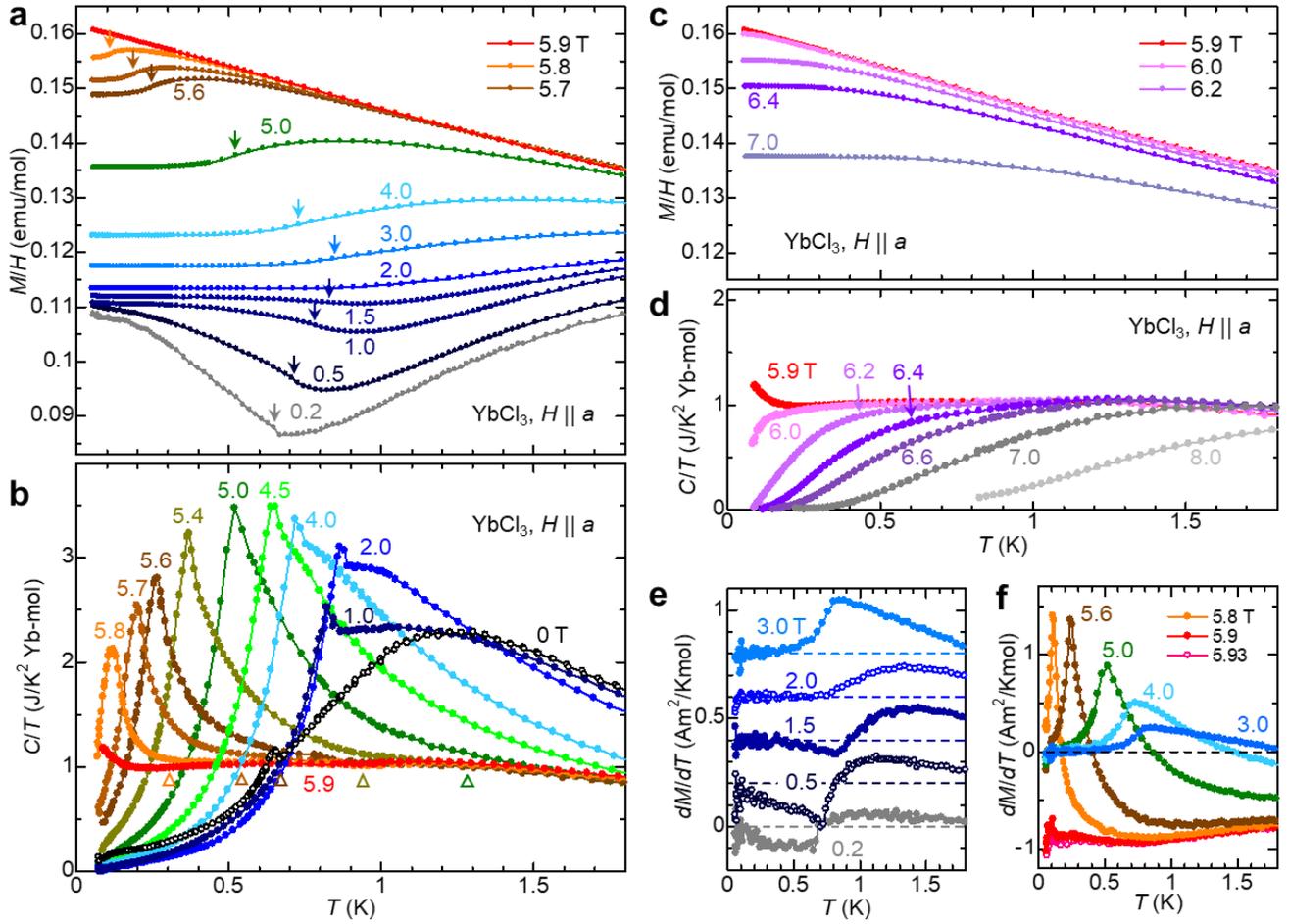

**Fig. 2 | Temperature dependence of magnetization *M(T)* and specific heat *C(T)* at different magnetic fields.** a, c) *T*-dependent *M/H* for $H \leq H_s$ and for $H \geq H_s$, respectively. The arrows in a) represents $T_c$ determined from the singularity in *dM/dT* (see **SI**). b, d) *T*-dependent *C/T* for $H \leq H_s$ and for $H \geq H_s$, respectively. Open triangles in b) mark the onset temperature for thermal fluctuations toward $T_c$, where *C/T* starts to deviate from the constant behavior upon cooling. e, f) *T*-dependence of the derivative of magnetization *dM/dT* below and above 3T, respectively. The data at each field in e) are shifted by 0.2 Am$^2$/Kmol for clarity.

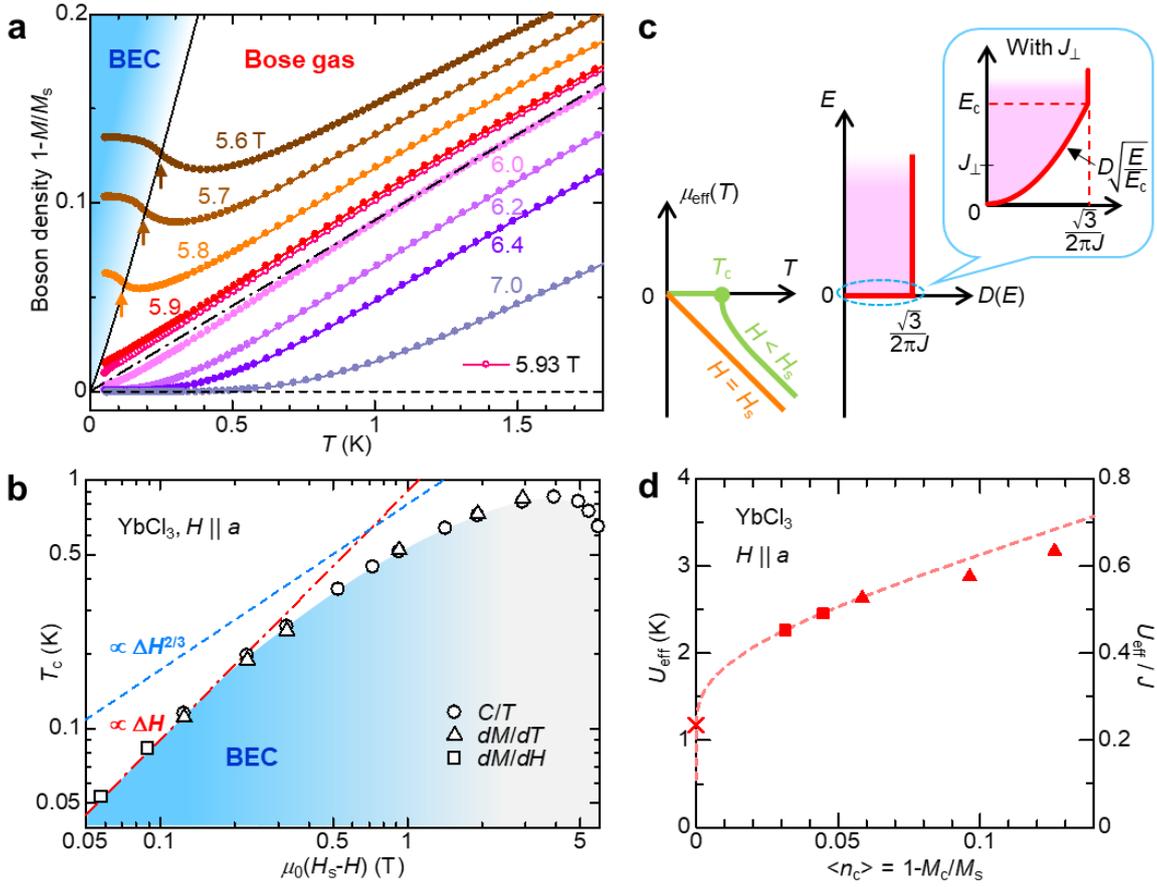

**Fig. 3 | Quantum critical behavior of magnetization $M(T)$ and $T_c(H)$, and Quasi-2D Bose gas model for analysis.** a) Temperature dependence of boson density $\langle n \rangle$ determined from the measured magnetization $M$ and the saturation moment $M_s$ as $\langle n \rangle = 1 - M/M_s$ for various magnetic fields $H$ around the QCP. The dashed line represents the linear behavior with $\langle n \rangle = T/T_0$ and $T_0 = 11$ K $= 2.2J$. The arrows indicate the magnetic transition temperature $T_c$ determined from the peak in $dM/dT$ (Figs. 2e, f). The solid line represents the relation between the critical density $\langle n_c \rangle$ and the BEC transition temperature $T_{BEC}$, $\langle n_c \rangle = Dk_B T_{BEC}(-\ln(k_B T_{BEC}/E_c) + 2)$ with a cut-off energy $E_c = 0.2$ mK and a density of states of $D = \sqrt{3}/2\pi J$ for $J = 5$K (See Fig. 3c). b) Full logarithmic plot of the scaling behavior of $T_c$ as a function of differential magnetic field from the QCP, $H_s - H$, with $H_s = 5.93$ T. The dashed and the broken lines represent $T_c \propto (H_s - H)$ and $(H_s - H)^{2/3}$ expected for 2D and 3D, respectively. c) The dilute Bose gas model used to analyze $C(T)$ and $M(T)$ around the QCP. The density of states $D(E)$ for the 2D honeycomb tight-binding model has a bandwidth of $3J$ and a finite value at $E = 0$, $D = \sqrt{3}/2\pi J$. In the presence of a small interlayer coupling $J_\perp$ ($\ll J$), a 3D density of states with $D(E)$ proportional to $\sqrt{E}$ shows up at the bottom of the band over the scale of $J_\perp$ and prevents the logarithmic divergence of the number integral in $T \to 0$ limit. We approximate this effect by introducing a cut-off of 2D constant density of states at $E_c$, below which $D(E)$ is replaced with $D\sqrt{E}/E_c$, with $E_c$ of the order of 2-3$J_\perp$. The effective chemical potential of the system is $\mu_{eff}(T) = g\mu_B(H_s - H) - 2U_{eff} \langle n \rangle$. At the QCP, $H = H_s$, $\mu_{eff}(T) = -2U_{eff}\langle n \rangle$ approaches to $E = 0$ linearly with $T$. For $H < H_s$, $\mu_{eff}$ goes to zero at a finite temperature due the suppression of the logarithmic divergence by the 3D DOS below $E_c$, where the boson density $\langle n \rangle = Dk_B T(-\ln(k_B T/E_c) + 2)$. This gives rise to a BEC. d) The effective boson-boson interaction $U_{eff}$ versus the critical boson density $\langle n_c \rangle$. $U_{eff}$ at $\langle n_c \rangle \to 0$ ($H = H_s$, cross) was estimated from the linear $T$-dependence of boson density $\langle n \rangle$ as described in the main text. For $H < H_s$, $U_{eff}$ was estimated from $\langle n_c \rangle$ at $T_c(H)$ in a) (triangles) and at $H_c(T)$ in Fig. S1c (squares) by using $U_{eff} = g\mu_B(H_s - H)/2\langle n_c \rangle$. The broken line represents the $\langle n_c \rangle$ dependence of $U_{eff}$ with 2D logarithmic renormalization and a cut-off by interlayer coupling, phenomenologically expressed by $U_{eff} = -U_0(1/\ln\langle n_c \rangle$

+ 1/ln($J_\perp/J$)). A bare boson-boson interaction $U_0 = 6$ K $= 1.2J$ and an interlayer coupling $J_\perp/J = 2 \times 10^{-5}$ are used.

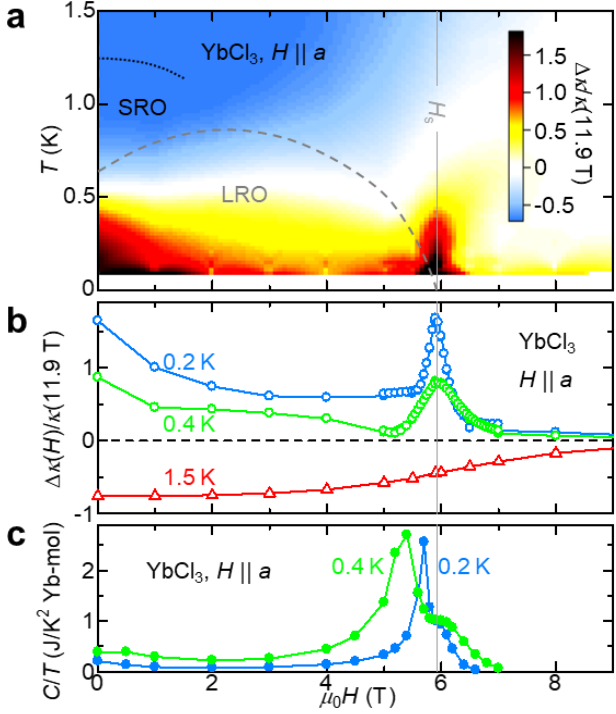

**Fig. 4 | Enhancement of thermal conductivity $\kappa$ at the quantum critical point $H_s$.** a) The normalized excess thermal conductivity $\Delta\kappa(H)/\kappa(11.9\ T) = [\kappa(T, H) - \kappa(T, 11.9\ T)]/\kappa(T, 11.9\ T)$ plotted as a color map on the magnetic field $H$ and the temperature $T$ plane. The excess conductivity $\Delta\kappa(H) = \kappa(H) - \kappa(11.9\ T)$ is the deviation from $\kappa(T)$ at 11.9 T. $\kappa(11.9\ T)$ data can be regarded as a phonon-only contribution to $\kappa$ in the absence of scattering by magnetic excitations. b) The $H$-dependence of $\Delta\kappa(H)/\kappa(11.9\ T)$ at 0.2, 0.4 and 1.5 K. c) The $H$-dependence of $C(H)/T$ at 0.2 and 0.4 K.

# Supplementary Information for "Quantum critical Bose gas in the two-dimensional limit in the honeycomb antiferromagnet YbCl₃ under magnetic fields" by Y. Matsumoto, S. Schnierer, J. A. N. Bruin, J. Nuss, P. Reiss, G. Jackeli, K. Kitagawa, and H. Takagi

## Magnetic field induced XY antiferromagnetism and BEC manifested in dimer-based systems and Heisenberg antiferromagnets

Magnon-BEC systems previously studied are mostly those based on a 3D network of dimers. At low fields, the dimer systems are gapped. With increasing $H$ to a lower critical field $H_{c1}$, a field-induced canted XY antiferromagnetism with a finite magnetization $<S_z>$ shows up. Eventually at a higher critical field, $H_{c2}$, the system is fully polarized and the canted XY AFM disappears. The canted XY-AFM between $H_{c1}$ and $H_{c2}$ is described as BEC. It is essentially the same physics as those for Heisenberg systems under magnetic fields described in the main text. The only difference is that the gapped region around $H = 0$ in dimer-based system is replaced with the Heisenberg AFM. Heisenberg-XY crossover emerges in Heisenberg system instead of $H_{c1}$ transition in dimer system. $H_s$ in Heisenberg system is equivalent to $H_{c2}$ in dimer systems.

## Determination of the saturation magnetization $M_s$, the Van Vleck susceptibility $\chi_{VV}$, the saturation field $H_s$ and the transition field $H_c$ from the $M$-$H$ curve

The saturation magnetization $M_s$ and the Van Vleck contribution to the magnetization $M$ was determined by a linear fit to the magnetization $M = M_s + \chi_{VV}H$, above $H_s$, which is shown by the gray dashed line in Fig. S1a. This yields the saturation magnetization $M_s = 1.72$ $\mu_B$/Yb and the Van-Vleck susceptibility $\chi_{VV} = 8.18 \times 10^{-3}$ emu/mol-Yb. The saturation field $\mu_0H_s = 5.93$ T was determined as the crossing point of $dM/dH$ at 0.05 and 0.08 K as shown in the inset to Fig. S1b. Given the general expression of $H_s$ for the nearest-neighbor Heisenberg antiferromagnet on a bipartite lattice, $H_s = 2zJS/g\mu_B$ with a coordination number $z$ and spin $S$, one can estimate the in-plane $g$ factor to be $g \sim 3.67$ using $J \approx 0.42$ meV and $\mu_0H_s = 5.93$ T. This gives $M_s \sim 1.84$ $\mu_B$/Yb, which is roughly consistent with 1.72 $\mu_B$/Yb. We note that the in-plane $g$ factor is consistent with $g = 3.6(1)$ estimated previously from the Curie-Weiss fit to $M/H$ [1]. To see the critical behavior of $T_c(H_c)$ near $H_s$ in Fig. 3b in the main text, we determined $H_c$ as the peak field of $dM/dH$ curve (Fig. S1b) at $T = 0.05$ K and 0.08 K. These two points are plotted as the open squares in Fig. 3b. The critical boson density $<n_c>$ at the respective temperature can be obtained from plot of $<n>(H) = 1 - M/M_s$ in Fig. S1c, which follows reasonably the $<n_c>(T)$ fitting curve in Fig. 3a of the main text.

## Subtraction of nuclear contributions from the measured specific heat $C(T)$

Figure S2 presents the temperature dependence of the total specific heat $C_{tot}$ in full logarithmic scale. A power-law behavior $C_{tot} \propto T^{-2}$ (or $C_{tot}/T \propto T^{-3}$), which originates from the nuclear contribution, is clearly seen at low temperatures for all the applied fields including zero field. Since the nuclear electric quadrupole contributions are negligibly small, the large nuclear specific heat at zero field should be due to the nuclear Zeeman contribution caused by a large internal field, which originates from the magnetic ordering of Yb moments. Note that the phonon contribution estimated from a non-magnetic analog LuCl₃ is negligibly small below a few K [1].

To determine the nuclear contribution, we fit the data with $C_{tot} = A_nT^{-2} + A_\alpha T^\alpha$ in a field range of $0 \leq \mu_0H \leq 5.6$ T, assuming that $C_{tot}$ is well described by a sum of the nuclear contribution ($\propto T^{-2}$) and a low-lying power-law contribution from Yb moments ($\propto T^\alpha$). As seen in Figs. S2a, 2b and 2c, the fitting works very well.

As shown in Figure S3a, $\alpha$ takes an almost constant value close to 2.5 at 0.5-5.6 T, except for zero-field, where it takes a significantly smaller value close to 1.5. At $\mu_0 H$ = 5.7 and 5.8 T, the electronic contribution overlaps considerably with the nuclear contribution because of the suppression of the ordering. Since the $T^\alpha$ dependence is masked by the nuclear contribution, it is difficult to determine $\alpha$ as a free parameter. Therefore, at these fields, we fixed $\alpha$ to 2.5 which was obtained at lower fields. At fields around the QCP ($H_s$), 5.9 and 6.0 T, it is even more difficult to separate the nuclear and electronic contributions. Therefore, for these fields, $A_n$ was determined by interpolating the values obtained at other fields close to $H_s$. The field dependence of $A_n$ is shown in Fig. S3b. $A_n$ is related to the local Yb moment $\boldsymbol{m}_{Yb}$ as $A_n = (\Lambda_{n,Yb}/\mu_0)|\mu_0 \boldsymbol{H} + A_{hf}\boldsymbol{m}_{Yb}|^2 + \Lambda_{n,Cl}\mu_0 H^2$. Here, $\Lambda_{n,Yb}$ and $\Lambda_{n,Cl}$ are the molar nuclear Curie constants for Yb and Cl. $A_{hf}$ = 102 T / $\mu_B$ is the hyperfine coupling constant for Yb. $A_n$ at $H = 0$ yields $|\boldsymbol{m}_{Yb}|$ = 0.99 $\mu_B$/Yb, which is in good agreement with the values of 1.06(4) $\mu_B$ [2], 0.8(1) $\mu_B$ [1] and 0.86(3) $\mu_B$ [3] obtained from neutron diffraction measurements. Above $H_s$, where $\boldsymbol{H}$ and $\boldsymbol{m}_{Yb}$ are parallel with each other, $|\boldsymbol{m}_{Yb}|$ ~1.6 $\mu_B$/Yb is estimated from $A_n$, in reasonable agreement with the saturation moment of 1.72 $\mu_B$/Yb from magnetization curve in Fig. 1a.

At fields above $H_s$, $C_{tot}$ is well fitted to $C_{tot} = A_n T^{-2} + A_g \exp(-\Delta/T)$, corresponding to a gap opening in the electronic contribution (Fig. S2c). As mentioned in the main text, the obtained gap $\Delta$ roughly agrees with those estimated from the boson density $<n> = 1 - M/M_s$ and is proportional to $H - H_s$.

**Excitation gap above $H_s$ estimated from the magnetization $M$ and the specific heat $C$**

Thermally activated behaviors of the boson density $<n> = 1-M/M_s$ and the specific heat $C$ at magnetic fields above $H_s$ are clearly seen in Figs. S4a, b. The broken lines are the fits to $<n> \propto \exp(-\Delta_n/T)$ and $C \propto \exp(-\Delta_C/T)$. As shown in Fig. S4c, $\Delta_C$ ~ 0.8 $\Delta_n$ is linear in $H-H_s$ and $k_B \Delta_C$ ~ 0.8 $g\mu_B (H-H_s)$. Because $\Delta_C$ and $\Delta_n$ are close to and scaled with each other, $\Delta_C$ is plotted in Fig. 1c as the magnetic excitation energy $\Delta$.

**Estimation of Entropy**

The temperature dependence of the electronic entropy $S(T)$ was determined by integrating $C/T$ at each field after the subtraction of the nuclear contribution. The contributions below the lowest temperature of the measurements were estimated by extrapolating the power-law behavior below $H_s$ or the gapped behavior above $H_s$, which were determined above. This offset is as small as 0.02-1% of $R\ln 2$. Only for 5.9 and 6.0 T near $H_s$ it was difficult to conduct such extrapolation due to the nontrivial quantum critical behavior of electronic part. Therefore, we determined the offset for these two fields so that these data smoothly interpolate the isothermal entropy at 1.4-1.6 K. The obtained isothermal entropy is shown in Fig. 1b in the main text. The obtained temperature dependence of entropy is summarized in Figure S5. At the QCP, the accumulation of entropy is clearly seen.

**Quantum critical region identified in $dM/dT$ map**

Figure S7 shows a map of $dM/dT$ plotted on the $H$-$T$ phase plane. A fan-shaped quantum critical region colored in red is identified, where the magnetization shows a quantum critical $T$-linear decrease and a negative and constant $dM/dT$. The quantum critical region from the magnetization coincides roughly that from the specific heat shown in Fig. 1c of the main text.

**Temperature dependence of thermal conductivity $\kappa$**

Figure S7a shows the temperature dependence of $\kappa/T$. As discussed in the main text, $\kappa(T)$ at 11.9 T serves as a reference for the pure phonon thermal conductivity without the magnetic scattering below 2 K. The magnetic excitations can be ignored below 2 K at 11.9 T as the excitation gap is much larger than 2 K. $\kappa(T)$

at high temperatures above ~ 0.6 K decreases monotonically with lowering $H$ from 11.9 T, reflecting the phonon-dominated thermal transport and the increase of magnetic excitations to scatter phonons due to the closure of the magnetic excitation gap at $H_s$. At lower temperatures below ~ 0.5 K, $\kappa(T)$ larger than that at 11.9 T, which indicates the emergence of additional magnetic contribution to $\kappa$.

Figure S8b presents the temperature dependence of the derivative of $\kappa$, $d\kappa/dT$. At zero-field, $d\kappa/dT$ shows a sharp dip at $T_c$. Upon application of field, the anomaly at $T_c$ is rapidly suppressed and impossible to identify above 4 T.

At high fields well above $H_s$, $d\kappa/dT$ shows a power-law behavior very close to $d\kappa/dT \propto T^2$ at low temperatures, which is indicative of the dominant phonon contribution at the low temperatures. Indeed, the phonon mean free path at 11.9 T reaches $l_{ph}$ ~ 60 µm in the low $T$ limit, which is estimated by using $C$ data from the non-magnetic analogue LuCl$_3$ [1] as the phonon specific heat of YbCl$_3$ and a phonon velocity of 2000 m/s. The long $l_{ph}$ corresponds to the "3D-averaged" sample dimension (1000 µm × 2000 µm × 20 µm)$^{1/3}$ ~ 340 µm, indicating that phonon transport is close to or even marginally in the ballistic regime. These confirm the high quality of the single crystals used in this study.

At the lower fields below 8 T, $d\kappa/dT$ in the low $T$ limit start to deviate from the $T^2$ behavior due to the magnetic contributions to $\kappa$ as well as the magnetic scattering of phonon contributions to $\kappa$. This change of power is clearly seen in the power-law exponent $\beta$ of the thermal conductivity, $\beta = d\ln\kappa/d\ln T$ ($\kappa \propto T^\beta$). Fig. S8 shows the color map of $\beta$. The inset shows the average below 0.25 K. $\beta$ is very close to 3 (corresponding to $T^2$ behavior of $d\kappa/dT$) above $H_s$ in the low $T$ limit and shows a lower value than 3 below $H_s$. We do see a sharp dip at $H_s$, reflecting the contribution from the highly mobile quantum critical Bose gas. Quantum critical Bose gas are discussed to show $\kappa \propto T^{1.5}$ in the 2D limit [4]. The spike-like decrease of power around $H_s$ is qualitatively consistent with the superposition of $T^{1.5}$ contribution. The presence of additional phonon contribution, however, does not allow us to isolate the magnetic contribution and to determine its power quantitatively.

### Density of states of 2D honeycomb boson band

The Matsubara-Matsuda transformation[5] turns the Hamiltonian of the nearest-neighbor Heisenberg spin into that of interacting bosons as in eq (1) in the main text. The kinetic part is a tight-binding Hamiltonian of bosons on the corresponding lattice with a hopping amplitude $t = J/2$. For 2D honeycomb lattice relevant to the case for YbCl$_3$, the tight binding Hamiltonian gives the following dispersion relation:

$$E(k) = \frac{3}{2}J \pm \frac{J}{2}\sqrt{1 + 4\cos\left(\frac{3}{2}k_x a\right)\cos\left(\frac{\sqrt{3}}{2}k_y a\right) + 4\cos^2\left(\frac{\sqrt{3}}{2}k_y a\right)}$$

Here, $a$ is the lattice constant. By expanding the dispersion around $k$ ~ 0, at the bottom of the band, we obtain $E(k)$ ~ $(3/8)Jk^2a^2$. The density of states per site at the bottom on band is calculated as

$$D = \frac{\sqrt{3}}{2\pi J}.$$

### Calculation of specific heat for a dilute-limit 2D boson gas

We consider 2D bosons in the dilute limit at $H_s$ with a constant density of states $D$ per site and an effective chemical potential $\mu_{eff} = -2U_{eff}\langle n\rangle$. When $\mu_{eff} \ll k_B T$, the internal energy is approximated as

$$E(T) = \int_0^\infty \frac{DE}{e^{(E-\mu_{eff})/k_B T} - 1} dE \cong \frac{\pi^2}{6} D(k_B T)^2 - 2U_{eff}(Dk_B T\langle n\rangle + \langle n\rangle^2) = \frac{\pi^2}{6} D(k_B T)^2 - 2U_{eff}T^2\left(\frac{Dk_B}{T_0} + \frac{1}{T_0^2}\right).$$

In the last equation, we used $\langle n \rangle = T/T_0$. The first term represents $E$ from free bosons ($\mu_{eff} = U_{eff} = 0$). The last two terms represent a correction due to the repulsive interaction $U_{eff}$. This yields

$$\gamma = \frac{C(T)}{T} = \frac{\pi^2}{3}k_B^2 D - 4U_{eff}\left(\frac{Dk_B}{T_0} + \frac{1}{T_0^2}\right).$$

Using $U_{eff} = 1.2$ K, $T_0 = 11$ K and $D = \sqrt{3}/2\pi J$ with $J = 5$ K, we find that $\gamma_0 = \frac{\pi^2}{3}k_B^2 D \sim 1.5$ J/K$^2$mol for free bosons is reduced to $\gamma \sim 0.99$ J/K$^2$mol due to $U_{eff}$. The estimated $\gamma$ is in excellent agreement with the experimental data.

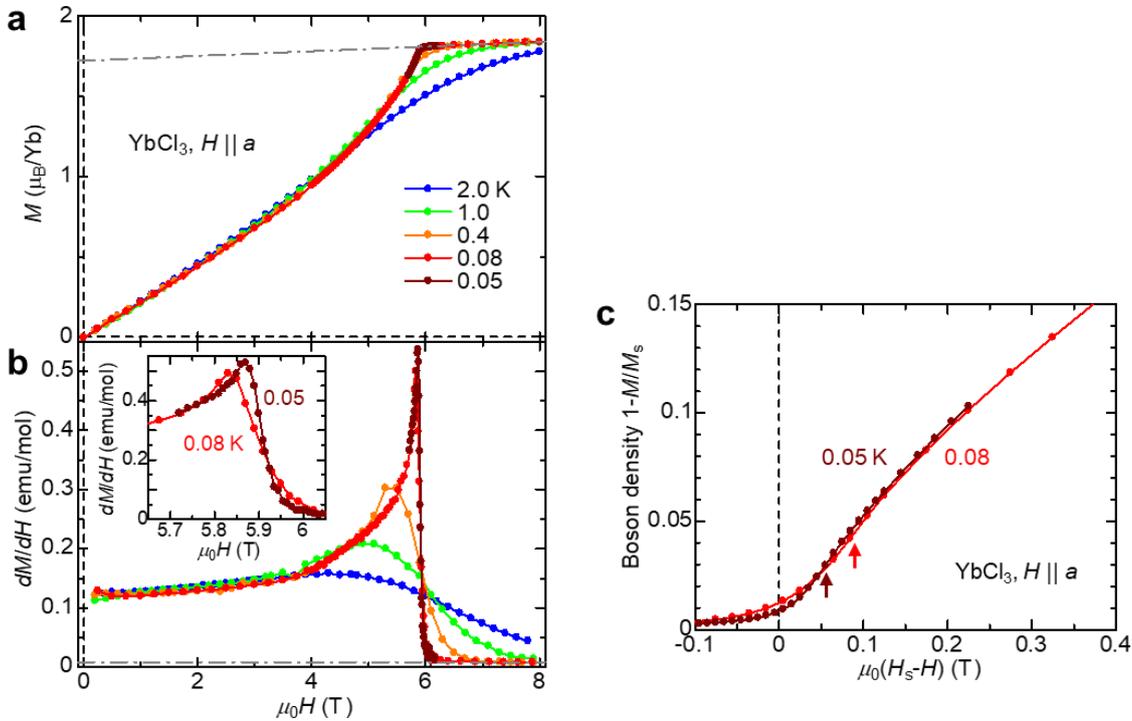

**Fig. S1. Magnetization process of YbCl$_3$ for magnetic field along the $a$-axis.** a) Magnetization curve $M(H)$ at various temperatures. The grey dashed line indicates the fit of the Van Vleck term. b) Field derivative of the magnetization $dM/dH$ as a function H at various temperatures The grey dashed line corresponds to the Van Vleck term. The inset shows $dM/dH$ at 0.05 and 0.08 K around $H_s$. c) Boson density $\langle n \rangle = 1 - M/M_s$ at the lowest temperatures as a function of differential magnetic field from the QCP, $H_s - H$, with $H_s = 5.93$ T. Arrows indicates the location of peak in $dM/dH$ shown in the inset for b).

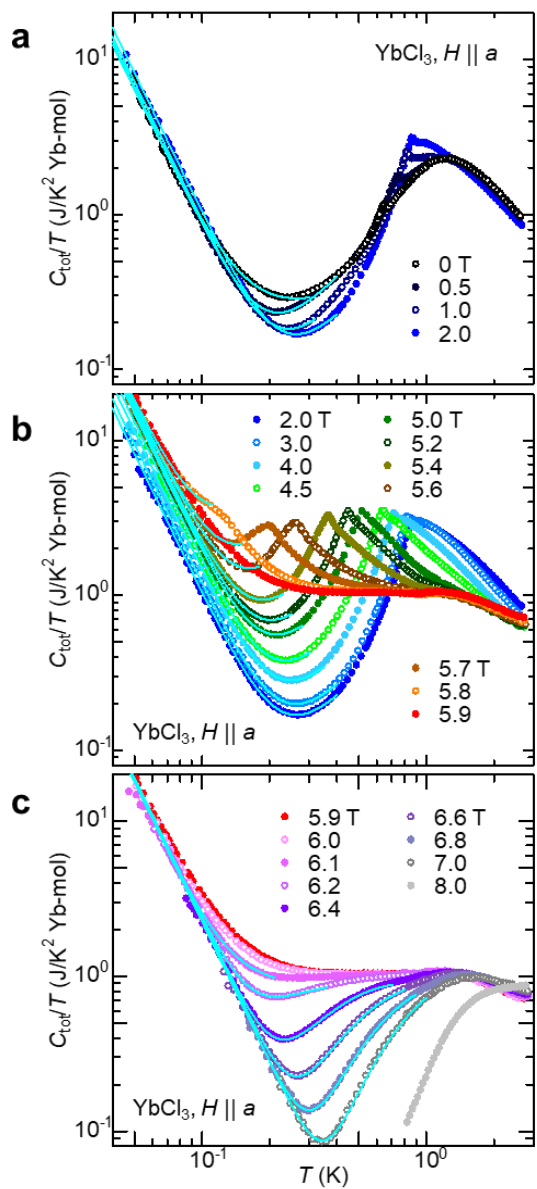

**Fig. S2. Temperature dependence of the total specific heat $C_{tot}$.** At a) low fields, b) intermediate fields below $H_s$ and c) high fields above $H_s$. The solid lines indicate the result of fitting $C_{tot} = A_n T^{-2} + A_\alpha T^\alpha$. See supplementary texts for details. The fitting allows us to subtract the nuclear contribution $A_n T^{-2}$ from $C_{tot}$.

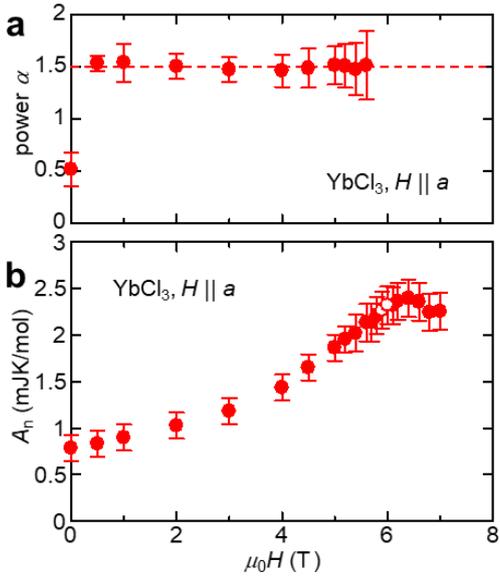

**Fig. S3. Field dependence of the parameters used to determine the nuclear contribution.** a) The power-law exponent $\alpha$ of the electronic specific heat and b) The nuclear Schottky contribution $A_n$ obtained from the fitting of $C_{tot} = A_n T^{-2} + A_\alpha T^\alpha$. Error bars represent one standard error estimated by varying the temperature range of the fitting by 10-20%. Open symbols are those for 5.9 and 6.0 T obtained by interpolation. See supplementary texts for details.

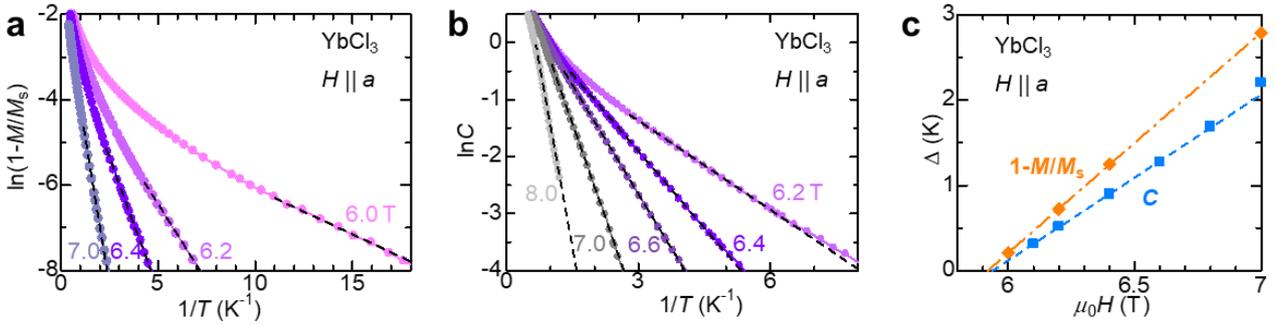

**Fig. S4. Arrhenius plot of the boson density $1-M/M_s$ and specific heat $C$ in fields above $H_s$.** a, b) The magnetic excitation energies $\Delta_n$ and $\Delta_C$ are extracted from the activated temperature dependence of boson density $<n>=1-M/M_s \propto \exp(-\Delta_n/T)$ in a) and $C \propto \exp(-\Delta_C/T)$ in b), as indicated by the dotted lines. c) Magnetic field dependence of excitation energies $\Delta_n$ and $\Delta_C$ obtained from the boson density $<n>$ and $C$ respectively. The lines represent the linear fits, where $\Delta_C \sim 0.8\Delta_n \sim 0.8g\mu_B(H-H_S)/k_B$.

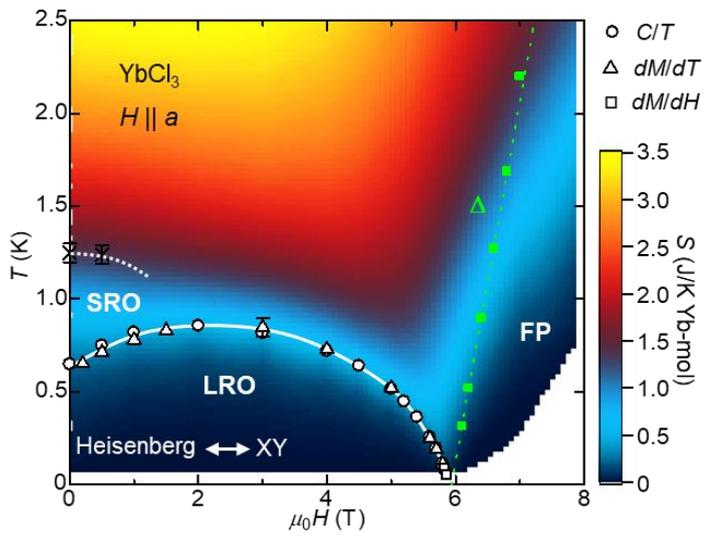

**Fig. S5. Entropy $S(T, H)$ map on the $H$-$T$ plane.** The symbols and the phase lines are the same as those in Fig. 1c.

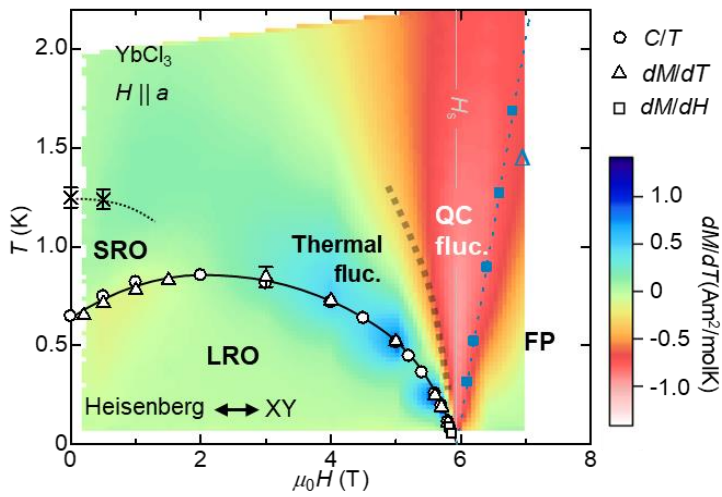

**Fig. S6. Map of temperature derivative of magnetization $dM/dT$ $(T,H)$ on the $H$-$T$ plane.** The symbols and the phase lines are the same as those in Fig. 1c.

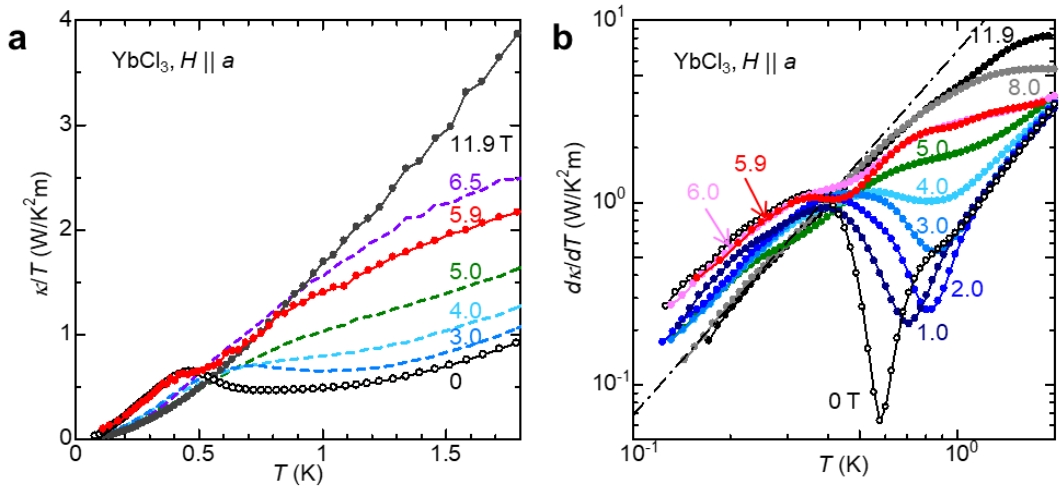

**Fig. S7. Temperature dependence of thermal conductivity $\kappa/T$ and its temperature derivative $d\kappa/dT$ for different magnetic fields.** a) $\kappa/T$ versus $T$ in linear scale. b) $d\kappa/dT$ versus $T$ in full logarithmic scale. The dashed line represents a power-law behavior $d\kappa/dT \propto T^2$.

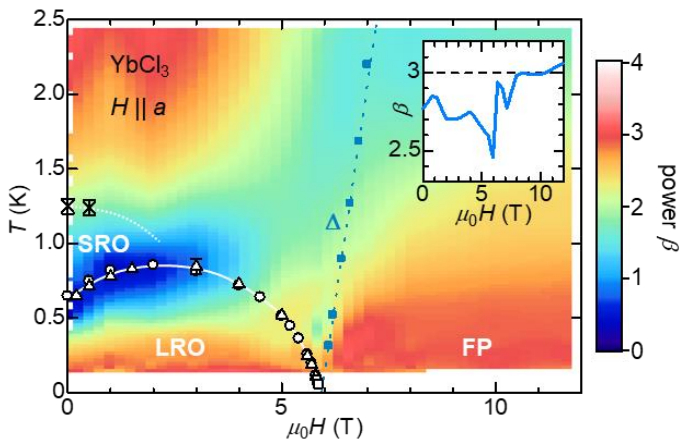

**Fig. S8. Color contour plot of the $T$-dependent power-law exponent $\beta(T) \equiv d\ln\kappa/d\ln T$ on top of the $H$-$T$ phase diagram.** The symbols are the same as Fig. 1b in the main text. Inset shows $\beta$ averaged below 0.25 K.